\documentclass[aps,prb,twocolumn,superscriptaddress,floatfix]{revtex4}%
\bibpunct{[}{]}{,}{n}{}{}
\usepackage{graphicx}
\usepackage{dcolumn}
\usepackage{latexsym}
\usepackage{amssymb}
\usepackage{amsfonts}
\usepackage{amsmath}
\usepackage{fancybox}
\usepackage[binary,squaren]{SIunits}
\usepackage{colordvi}
\usepackage{color}%
\allowdisplaybreaks[4]
\providecommand{\U}[1]{\protect\rule{.1in}{.1in}}
\providecommand{\U}[1]{\protect\rule{.1in}{.1in}}

\def\be{\begin{equation}}
\def\ee{\end{equation}}
\def\ber{\begin{eqnarray}}
\def\eer{\end{eqnarray}}

\def\taubold{\mbox{\boldmath $\tau$}}

\begin{document}
\title{Theory of spin and lattice wave dynamics excited by focused laser pulses}
\author{Ka Shen}
\email{kashen@bnu.edu.cn}
\affiliation{The Center for Advanced Quantum Studies and Department of Physics, Beijing Normal University, Beijing 100875, China}

\affiliation{Kavli Institute of NanoScience, Delft University of Technology, Lorentzweg 1,
2628 CJ Delft, The Netherlands}

\author{Gerrit E. W. Bauer}
\affiliation{Institute for Materials Research and WPI-AIMR, Tohoku University, Sendai
980-8577, Japan}
\affiliation{Kavli Institute of NanoScience, Delft University of Technology, Lorentzweg 1,
2628 CJ Delft, The Netherlands}
%\date{\today }

\begin{abstract}
We develop a theory of the spin wave dynamics excited by ultrafast focused laser pulses in a magnetic film. We take into account both volume and surface spin wave modes in the presence of applied, dipolar and magnetic anisotropy fields and include the dependence on laser spot exposure size and magnetic damping. We show that the sound waves generated by local heating by an ultrafast focused laser pulse can excite a wide spectrum of spin waves (on top of a dominant magnon-phonon contribution). Good agreement with recent experiments supports the validity of the model.
\end{abstract}

\maketitle
% Uncomment for PACS numbers
%\pacs{75.80.+q, 75.30.Ds, 75.78.-n, 78.20.Ls}
%
% Uncomment for keywords
%\vspace{2pc}
%\noindent{\it Keywords}: spin waves dynamics, pump-and-probe, magnons, phonons, magnetoelasticity

% Uncomment for Submitted to journal title message
%\submitto{\JPD}
%
% Uncomment if a separate title page is required
%\maketitle
% 
% For two-column output uncomment the next line and choose [10pt] rather than [12pt] in the \documentclass declaration

%\ioptwocol
%

\section{Introduction}
Generation, manipulation, and detection of spin polarization by ultrafast laser pulses are powerful techniques in the field of spintronics and magnetism~\cite{Ju99,Kampen02,Vahaplar09,Kirilyuk10,Barker13}. Such operations rely on laser pulse-induced demagnetization due to heating~\cite{Beaurepaire96}, the change of magnetic anisotropy due to charge transfer~\cite{Ju99,Kampen02}, magneto-optical (inverse) Faraday effect (IFE)~\cite{Kimel05,Hansteen05,Satoh12}, and (inverse) Cotton-Mouton effect~\cite{BenAmer11}. Recent experiments demonstrated magnetoelastic coupling of coherent acoustic sound waves generated by the local heating caused by optical gratings~\cite{Janusonis16,Chang17} and laser spots~\cite{Au13, Ogawa15,Shen15,Hashimoto17} with spin waves. Here we further develop the theory~\cite{Shen15} to analyze the spin wave excitation and  subsequent dynamics for the latter scenario.

Laser-induced spin (wave) dynamics has recently been studied in a pump-probe setup, in which a strong pump pulse is applied to excite the system and a second pulse is used to read out the transient dynamics of the out-of-plane magnetization by the Kerr or Faraday rotation of linearly polarized light. In our previous work~\cite{Shen15}, we simulated the magnetization dynamics coupled with acoustic waves by the magnetoelastic interaction. We assumed that the film is thick enough to allow simplifications such as the disregard of surface waves. Here we extend our previous model by taking into account the dipolar volume and surface spin waves of magnetic films with arbitrary thickness~\cite{Damon1961}, but disregard surface effects on the phonons. 

The modeling of the coupled magnon-phonon system involves only a few and rather well-known parameters. We assume that the pump pulse generates heat. The associated sudden lattice expansion generates effective magnetic fields via the magnetoelastic coupling. Thereby, the local heating can excite spin waves not only in the reciprocal space regime in which spin wave and phonon frequencies are the same~\cite{Ogawa15,Shen15,Janusonis16,Kikkawa_polaron,Benedetta_polaron}, but also in the non-resonant regime as observed in recent experiments~\cite{Hashimoto17}. Spin-wave-phonon resonant excitation causes clear interference patterns in the the spatial dynamics, which are suppressed in the non-resonant case. The resonant via non-resonant excitation can be tuned by the magnetic field or the size of the laser spot. Moreover, the presence of large magnetic damping suppresses the interference features in resonant excitation. Our numerical results also reproduce the interference pattern in momentum space as observed in experiments~\cite{Hashimoto17}.

\section{Formalism}
\subsection{Damon-Eshbach theory}
We consider a magnetic film in the $x$-$y$ plane made from a cubic crystal with both cubic and uniaxial magnetic anisotropyies. The anisotropy energy reads
\begin{equation}
\mathcal {H}_{an}=K_c (m_x^2 m_y^2 +m_x^2 m_z^2 +m_y^2 m_z^2) +K_u (1-m_z^2),
\end{equation}
where we assume a cubic anisotropy coefficient $K_c>0$ and uniaxial anisotropy coefficient $K_u<0$. The equilibrium magnetization is aligned with an external magnetic field along the $x$-axis. For weak excitation the magnetization unit vector 
normalized by the saturation magnetization $M_0$
then reads  $\mathbf m (\mathbf r,t) \simeq(1, m_y, m_z)$ where $|m_{x/y}|\ll 1$. To leading order the anisotropy and Zeeman energies are
\begin{equation}
\mathcal{H}_{an}+\mathcal{H}_{Z} \simeq \frac{\mu_{0}}{2}{\tilde H}M_{0} m_y^2 + \frac{\mu_{0}}{2}{\tilde H}'M_{0} m_z^2 
\end{equation}
with $\tilde{H}=H+{2K_c}/({\mu_0 M_{0}})$ and $\tilde{H}'=\tilde{H}-{2K_u}/(\mu_0{M_{0}})$, where we omitted an irrelevant constant. The transverse magnetization components, $m_y$ and $m_z$ introduce an effective dipolar
field $\mathbf h (\mathbf r,t)$ in the equations of motion~\cite{Damon1961,Hurben96}
\begin{eqnarray}
-\omega^{2}m_{y}&=&-i\omega\gamma \mu_0 h_{z}+\gamma \mu_0 \tilde{H}'(\gamma \mu_{0}h_{y}-\gamma \mu_0 \tilde{H}m_{y}), \label{EOM_M} \\
-\omega^{2}m_{z}&=&i\omega\gamma \mu_{0}h_{y}-\gamma \mu_0\tilde{H}(-\gamma \mu_{0}h_{z}+\gamma \mu_0\tilde{H}'m_{z}). \label{EOM_Mz}
\end{eqnarray}
Here, $\gamma$ is the gyromagnetic ratio. We disregard the exchange coupling since it is small compared to the dipolar and anisotropy fields for the wave vector range of our present interest ($\sim 1/\mu$m)~\cite{Shen15,Hashimoto17}. 
 By applying continuity boundary conditions of in-plane $h_y$ and out-of-plane $h_z+M_0 m_z$ fields at the interfaces to the vacuum and non-magnetic substrate, we arrive at the characteristic equation~\cite{Damon1961,Hurben96}
\begin{eqnarray}
&&(1+\kappa)^{2}\left[(k_z^{(i)})^2/k^2\right]-2(1+\kappa)(k_z^{(i)}/k)\cot({k_{z}^{(i)}d})\nonumber \\ 
&&\hspace{1cm}=1-v^{2}\sin^{2}\theta \label{eigen_mode}
\end{eqnarray}
where 
\begin{eqnarray}
(k_{z}^{(i)})^2&=&-{\frac{1+\tilde{\kappa}\sin^{2}\theta}{1+\kappa}}k^{2}\\
\kappa&=&\frac{\omega_{H}\omega_{M}}{(\omega_{H}\omega_{H'}-\omega^{2})}\\
\tilde \kappa&=&\frac{\omega_{H^\prime}\omega_{M}}{(\omega_{H}\omega_{H'}-\omega^{2})}\\
\nu&=&\frac{\omega\omega_{M}}{(\omega_{H}\omega_{H'}-\omega^{2})}
\end{eqnarray}
and $d$ and $\theta$ are the film thickness and the angle between in-plane wave vector $\mathbf k$ and $x$-axis, respectively. Here, $\omega_{M}=\gamma\mu_{0}M_{0}$, $\omega_{H}=\gamma\mu_{0}{\tilde H}$, and $\omega_{H'}=\gamma\mu_{0}{\tilde H}'$. The solutions of Eq.~(\ref{eigen_mode}) are the spin wave frequency dispersion $\omega_{j\mathbf k}$ for a magnon with wave vector $\mathbf k$ and band index $j$. The corresponding wave functions $({\cal M}_y(z), {\cal M}_z(z))_{j\mathbf k}$ are obtained by substituting the frequencies  into the boundary conditions of Eqs.~(\ref{EOM_M}) and (\ref{EOM_Mz})~\cite{Damon1961}.

\subsection{Linear response theory}
In linear response the spatiotemporal dynamics of the non-equilibrium magnetization driven by a transverse magnetic field reads
\begin{eqnarray}
m_{i}(z,\mathbf{r},t)&=&\frac{1}{M_0}\sum_{\mathbf k}e^{i\mathbf k\cdot\mathbf r}\nonumber\\
&&\hspace{-0.4cm}\times\int dz' dt^{\prime}
\chi_{ij}(z,z',\mathbf k,t,t')
H_T^j(z',\mathbf k,t'), \label{Dym}%
\end{eqnarray}
where $\chi$ and $H_T$ are the spin susceptibility tensor and total effective magnetic field, respectively, with $i,j=y,z$. 

We derive the multi-mode spin susceptibility by introducing the torque exerted by the  transverse field, $\taubold=\gamma \hat x\times  (\mu_0 \mathbf H_T)$, i.e.,
\begin{equation}
\left(\begin{array}{c}
\tau_{m_{y}}(z)\\
\tau_{m_{z}}(z)
\end{array}\right)(\mathbf{k},t)=\left(\begin{array}{c}
-\gamma \mu_0 H_T^z(z)\\
\gamma \mu_0 H_T^y(z)
\end{array}\right)(\mathbf{k},t),
\label{eq:degeneration_rate}
\end{equation} 
which can be expanded into the spin wave modes
\begin{equation}
\left(\begin{array}{c}
\tau_{m_{y}}(z)\\
\tau_{m_{z}}(z)
\end{array}\right)(\mathbf{k},t)
=\sum_{j}a_{j\mathbf{k}}(t)\left(\begin{array}{c}
{\cal M}_{y}(z)\\
{\cal M}_{z}(z)
\end{array}\right)_{j\mathbf k},
\end{equation}
where $a_{j\mathbf{k}}(t)$ is the excitation rate of the $j$-th mode. From the orthonormality condition~\cite{Verba12}
\begin{equation}
\int dz (\mathcal M_{y,j}^\ast\mathcal M_{z,j'}-\mathcal M_{z,j}^\ast\mathcal M_{y,j'})=i\delta_{jj'}, \label{orthN}
\end{equation}
we observe that the expansion coefficients are a mode-resolved Zeeman energy
\begin{equation}
a_{j\mathbf k}=-i\gamma \mu_0 \int dz (\mathcal M_{y,j}^\ast H_{T}^y+\mathcal M_{z,j}^\ast H_T^z).
\end{equation}

At a given time $t$, all spin waves generated in the past $t'<t$ can contribute
to the local magnetization. The precession of a spin wave mode  $j\mathbf k$ excited at time $t'$ carries out precessional motion and reads at time $t$
 \begin{equation}
\left(\begin{array}{c}
m_{y}(z)\\
m_{z}(z)
\end{array}\right)_{j\mathbf k}(t)	=	 
a_{j\mathbf{k}}(t')e^{-i\omega_{j\mathbf{k}}(t-t')}\left(\begin{array}{c}
{\cal M}_{y}(z)\\
{\cal M}_{z}(z)
\end{array}\right)_{j\mathbf{k}}.\label{Dym2}
\end{equation}
By substituting $a_{j\mathbf{k}}$, we arrive at an explicit expression for the spin susceptibility
\begin{eqnarray}
\chi(z,z',\mathbf k,t,t')&=&-i\gamma\mu_0 M_0 \sum_j e^{-i\omega_{j\mathbf k}(t-t')}\nonumber\\
&&\hspace{-2cm}\times
\left(\begin{array}{cc}
\mathcal M_{y,j}(z)\mathcal M_{y,j}^\ast(z') &\mathcal M_{y,j}(z)\mathcal M_{z,j}^\ast(z')\\
\mathcal M_{z,j}(z)\mathcal M_{y,j}^\ast(z') &\mathcal M_{z,j}(z)\mathcal M_{z,j}^\ast(z')
\end{array}\right)_{\mathbf k}.
\end{eqnarray}
Dissipation can be included by introducing a prefactor $(1-i\alpha)$ in front of $\omega_{j\mathbf k}$, with $\alpha$ being the Gilbert damping constant.

Next we derive the driving field $H_T$ in Eq.~(\ref{Dym}). In the long-wavelength limit, the magnetoelastic coupling energy~\cite{Kittel58,Akhiezer58} in our configuration reads
\begin{equation}
\mathcal{H}_{\mathrm{mec}}=\sum_{i,j\in\{y,z\}}(b+a\delta_{ij})S_{ij}%
m_{i}m_{j}+2b\sum_{i\in\{y,z\}}S_{ix}m_{i},\label{MEC_H}
\end{equation}
which gives rise to a spin generating torque 
\begin{eqnarray}
\tau_{m_y}&=&(b/\hbar)(S_{xz}+S_{zx}),\\
\tau_{m_z}&=&(b/\hbar)(S_{xy}+S_{yx}),
\end{eqnarray}
where $a$ and $b$ are magnetoelastic coupling coefficients and $S_{ij}=(\partial_i R_j +\partial_j R_i)/2$ are the stresses due to a lattice displacement $R_{x,y,z}$. With some algebra~\cite{Shen15,Schlomann60} and comparison with Eq.~(\ref{eq:degeneration_rate}), we find the resulting effective magnetic field acting on a spin wave with wave vector $\mathbf k$
\be
\mathbf H_{T}(z,\mathbf k,t) =i\Delta_1[\hat y (R_l\sin 2\theta +R_t\cos 2\theta )+\hat z  R_z \cos \theta ] \label{HT}
\ee
where $\Delta_1=b k/(\hbar \gamma \mu_0)$ and $R_l$, $R_t$ are the longitudinal and transverse displacements relative to $\mathbf k$. 

\subsection{Lattice displacement due to heating}
Here, we model the lattice distortion created by the heating effect of a focused laser. Since local thermalization is complete within several picoseconds~\cite{Zijlstra13} after the pump pulse, the local temperature rises step-like on the nanosecond scale of the magnetization dynamics. We therefore assume a temperature profile
\be
T(\mathbf{r},t)=T_0+\delta T\cdot\Theta(t)\exp(-r^{2}/W^{2}-\Gamma t),
\ee 
where $\delta T$ is the temperature increase at the center of the exposed area. $W$ and $\Gamma$ describe the spot size and temperature relaxation rate, respectively, while we disregard the slow heat diffusion~\cite{Ogawa15}. The sudden temperature increase is a thermodynamic force to which the lattice reacts by a rapid elastic deformation accompanied by emission of lattice waves in the form of, e.g., Rayleigh and surface skimming longitudinal waves~\cite{Janusonis16}, as well as bulk phonons. The spatial profile of the displacement dynamics can, in principle, be calculated numerically by a careful treatment of the boundaries, leading to a mode-dependent penetration depth. Here we simplify our task by disregarding the mode-dependent mixing between $R_z$ and $R_l$, which as we will show later retain all observed features in the experiments.

We introduce the penetration depth $d_p$
as a parameter, i.e., $R_z(z,\mathbf r,t)=e^{-z/d_p}R_z(\mathbf r,t)$ and $R_l(z,\mathbf r,t)=e^{-z/d_p}R_l(\mathbf r,t)$, where the displacements at the free surface $R_z(\mathbf r,t)$ and $R_l(\mathbf r,t)$ are obtained from the equations of motion,
\ber
\partial_{t}^{2}R_{l}(\mathbf{r},t)-c_{l}^{2}\nabla^{2}R_{l}(\mathbf{r},t)&=&\eta(3\lambda+2\mu)\partial_{r}T(\mathbf{r},t),
\label{eq:EOM_Rl}\\
\partial_{t}^{2}R_{z}(\mathbf{r},t)-c_{t}^{2}\nabla^{2}R_{z}(\mathbf{r},t)&=&\zeta\eta\nabla^{2}T(\mathbf{r},t).\label{eq:EOM_RZ}
\eer
Here, we introduce an in-plane thermoelastic pressure stress~\cite{Davies93,Rossignol05} and a \textquotedblleft bulge\textquotedblright\ shear stress~\cite{Dewhurst82}, respectively. $c_t$ and $c_l$ are transverse and longitudinal sound velocity. We introduced the thermoelastic expansion coefficient $\eta$ and a thickness-dependent parameter $\zeta$~\cite{Shen15}. Eq.~(\ref{eq:EOM_RZ}) has the static solution $R_z(\mathbf r)\propto \delta T(\mathbf r)$.

Eqs.~(\ref{eq:EOM_Rl}) and (\ref{eq:EOM_RZ}) can be solved in momentum space, leading to
\ber
 R_{l}(\mathbf{k},t)&=& C_1 k\int\frac{d\omega}{2\pi}\frac{\exp({i\omega t-W^{2}k^{2}/4})}{(\omega+c_{l}k)(\omega-c_{l}k)(\omega-i\Gamma)},\label{eq:Rl_kt}\\
 R_{z}(\mathbf{k},t)&=& C_2 k^{2}\int\frac{d\omega}{2\pi}\frac{\exp(i\omega t-W^{2}k^{2}/4)}{(\omega+c_{t}k)(\omega-c_{t}k)(\omega-i\Gamma)},\label{eq:Rz_kt}
\eer
with $C_1=\eta (3\lambda+2\mu) \delta T$ and $C_2=\zeta\eta \delta T$.  $R_l$ and $R_z$ are the sources of the laser-induced magnetoelastic torques. We assume that the back action of the magnetization on the lattice is a weak perturbation that may be disregarded.

\begin{figure}[ptb]
\includegraphics[width=3.2cm]{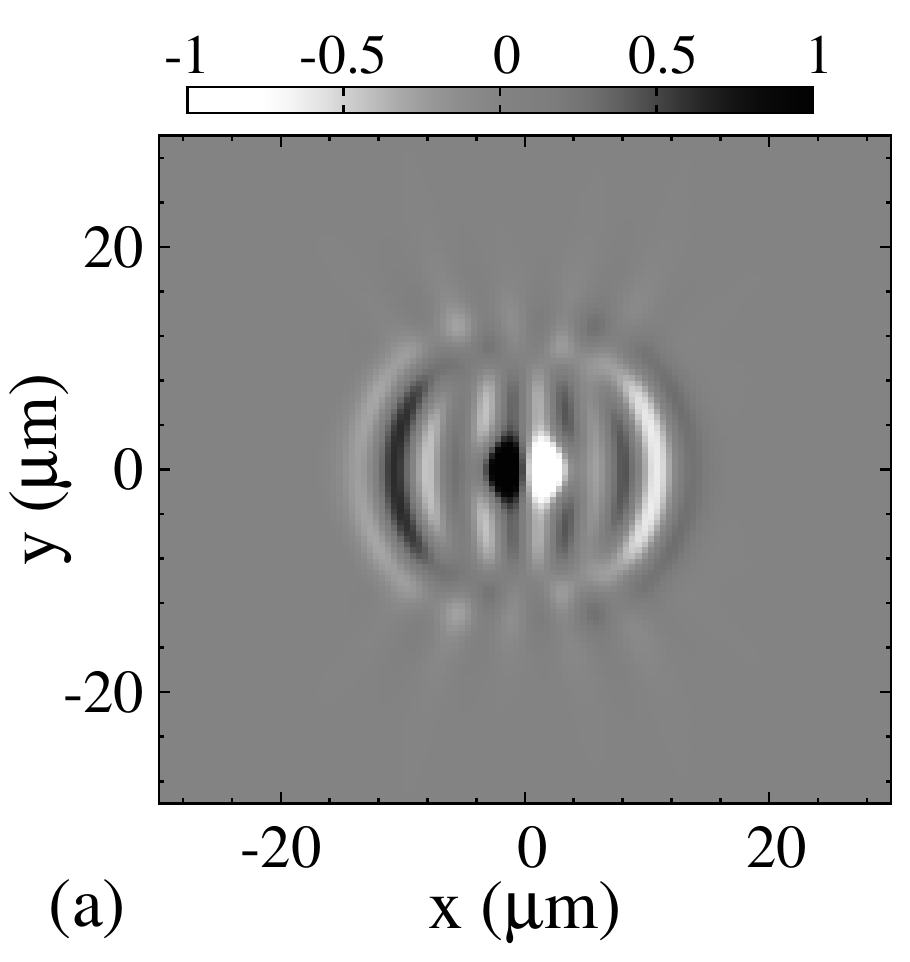}
\includegraphics[width=3.2cm]{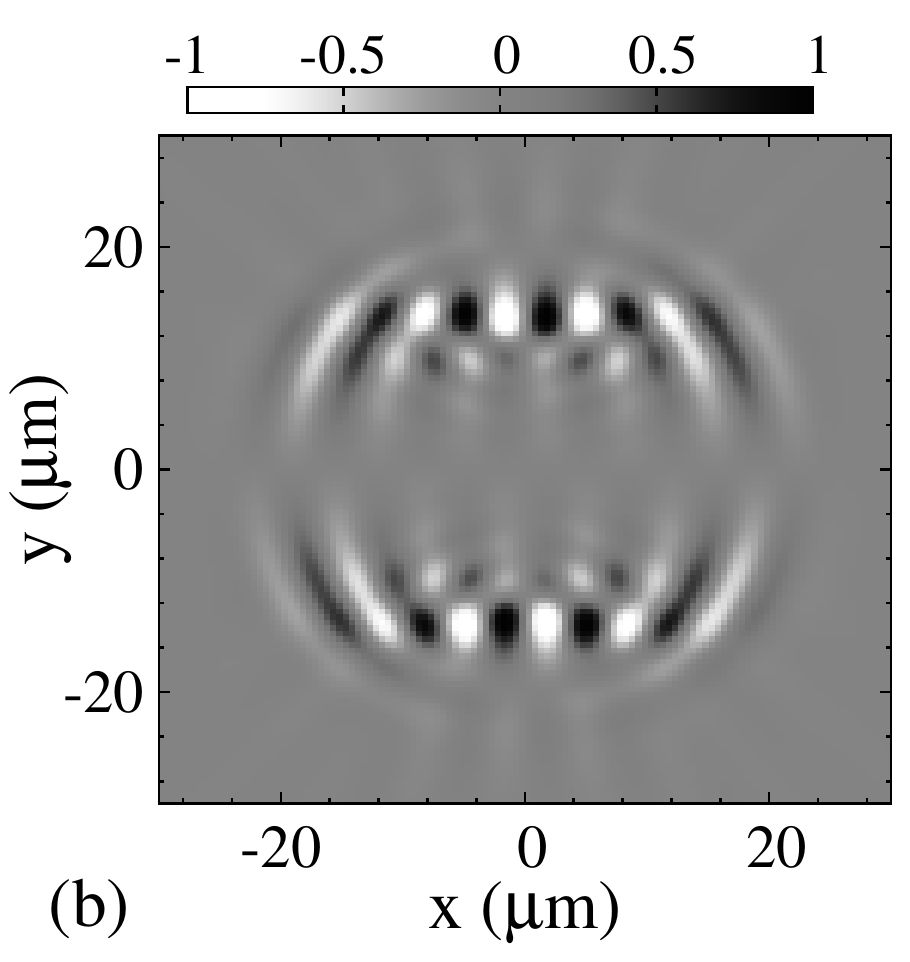}

\includegraphics[width=3.2cm]{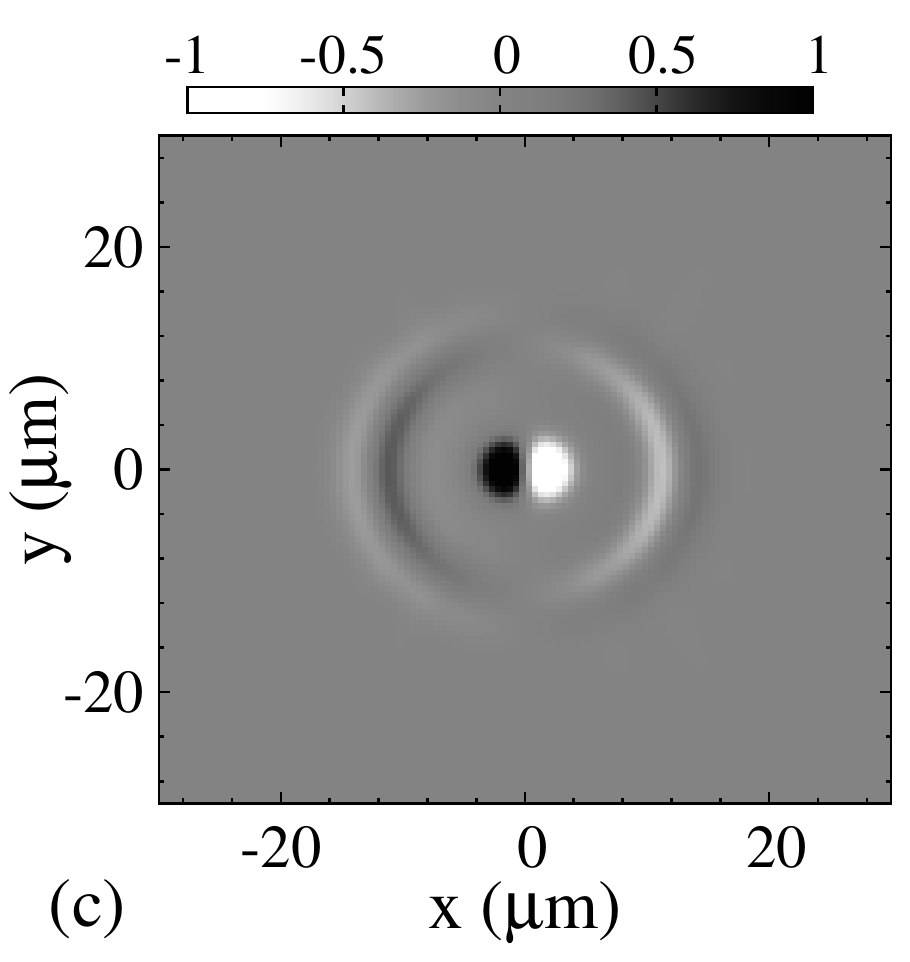}
\includegraphics[width=3.2cm]{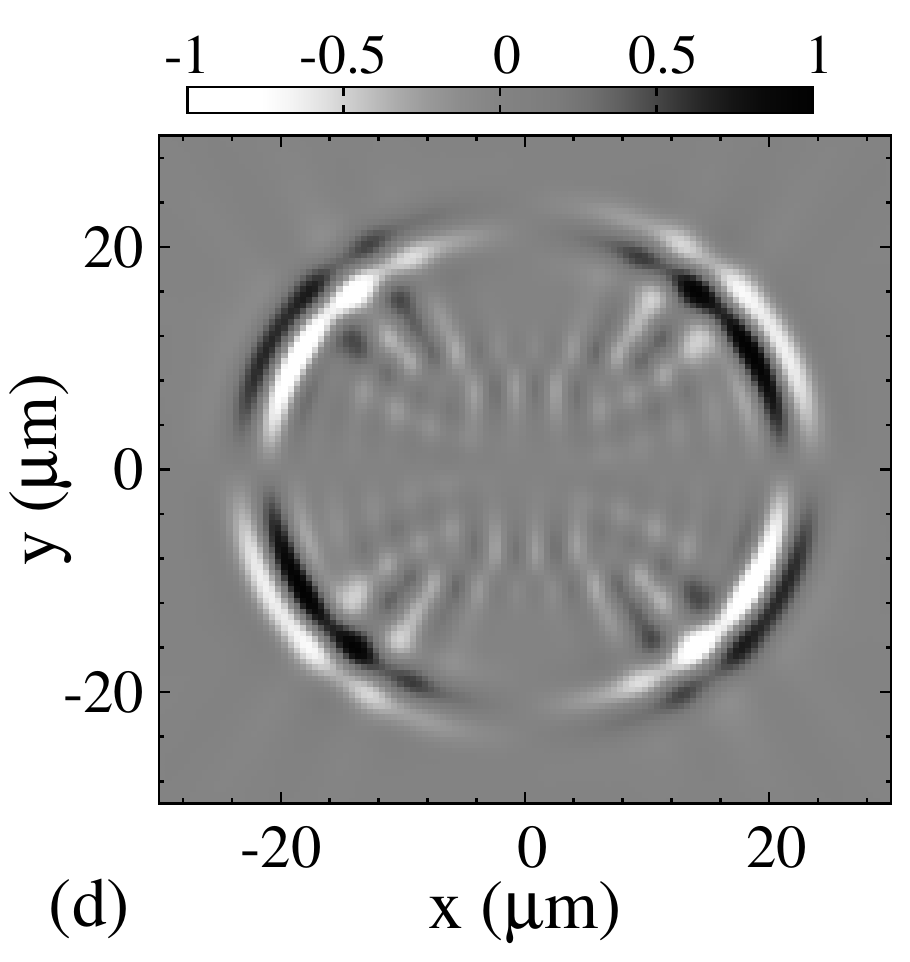}

\includegraphics[width=3.2cm]{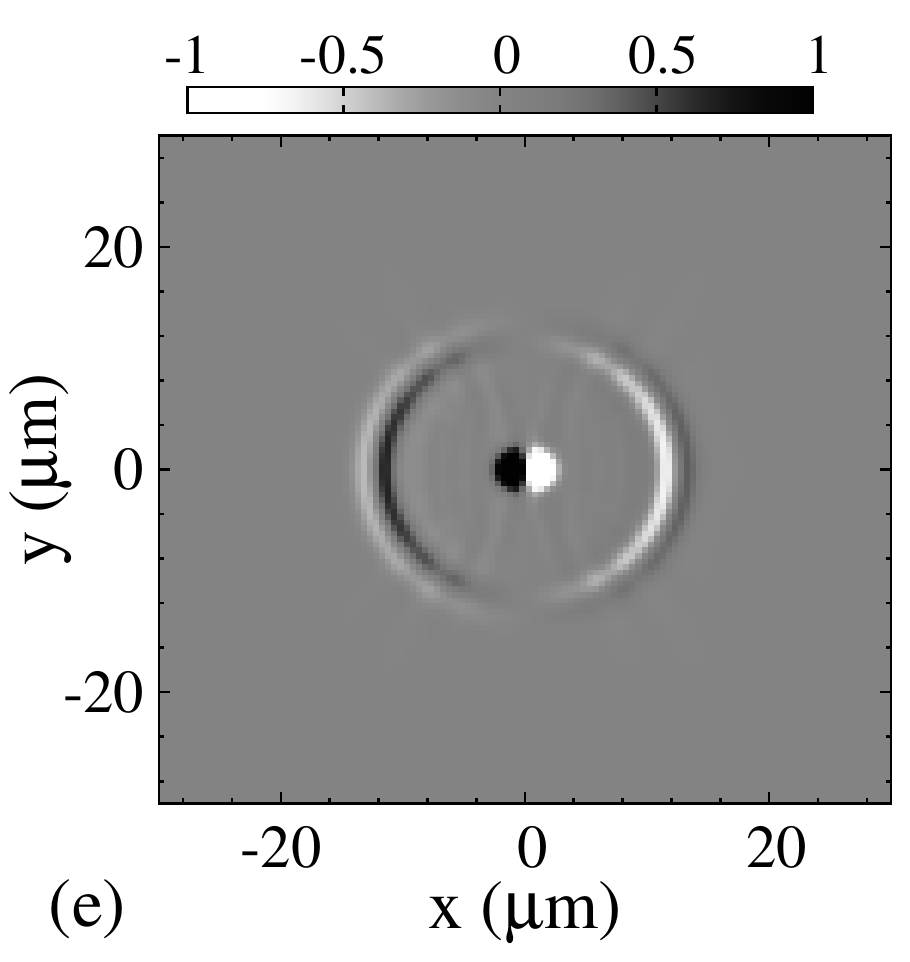}
\includegraphics[width=3.2cm]{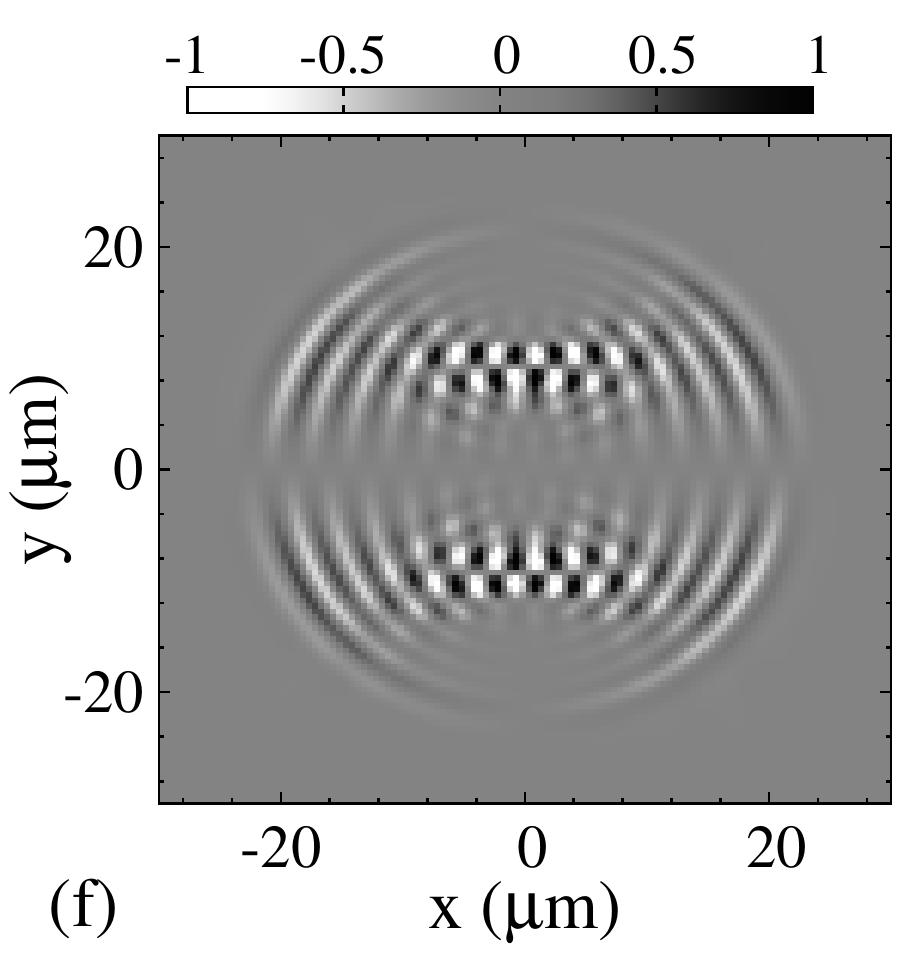}

\caption{Snapshot at 4~ns after an ultrafast laser pulse due to shear stress (left) and pressure stress (right) with external magnetic field and spot size in (a)-(b) $\mu_0 H=4$~mT and $W=2.5$~$\mu$m, (c)-(d) $\mu_0 H=40$~mT and $W=2.5$~$\mu$m, (e)-(f) $\mu_0 H=40$~mT and $W=1.5$~$\mu$m. The gray scale reflects the out-of-plane magnetization in arbitrary units. }%
\label{Mechanisms}%
%/Hashimoto/film/0for_paper/0with_anisotropy/demagnetization/osummary_mechanism.lyx
\end{figure}

\section{Results} 
Here we present numerical solutions to the system of equations formulated in the previous sections and compare them  with recent experiments. Assuming that the film is thinner than the optical extinction length, the Faraday/Kerr-type measurement, the Faraday/Kerr angle ($\Theta_{F}$) is proportional to the out-of-plane magnetization $m_{z}$ integrated over the magnetic film, i.e., 
\ber \Theta_{F}(\mathbf r,t) \propto \sum_{\mathbf k}e^{i\mathbf{k}\cdot\mathbf{r}}\sum_j\overline{ m_{z,j}(\mathbf{k},t)}.\label{eq:kerr}
\eer
with $\overline{ m_{z,j}(\mathbf{k},t)}=\int m_{z,j}(\mathbf{k},z,t) dz$. The heat diffusion is slow compared to spin wave/sound wave frequencies, i.e., $\Gamma\ll|\omega_{j\mathbf{k}}|,|c_{t}k|$, therefore
\ber
&&\overline{
 m_{z,j}(\mathbf{k},t)}\propto \overline{ e^{-z/d_p}\mathcal M^\ast_{z,j,\mathbf k}}\cdot \overline{ \mathcal M_{z,j,\mathbf k}} ke^{-W^{2}k^{2}/4} \cos\theta \nonumber\\
&&\times\left[\frac{e^{-i\omega_{j\mathbf{k}}t}-1}{\omega_{j\mathbf{k}}}-\frac{e^{-i\omega_{j\mathbf{k}}t}-e^{-ic_{t}kt}}{2(\omega_{j\mathbf{k}}-c_{t}k)}-\frac{e^{-i\omega_{j\mathbf{k}}t}-e^{ic_{t}kt}}{2(\omega_{j\mathbf{k}}+c_{t}k)}\right],\nonumber\\
\label{mzRkt}
\eer
from the shear and 
\ber
&&\overline{
 m_{z,j}(\mathbf{k},t)}\propto \overline{ e^{-z/d_p}\mathcal M^\ast_{y,j,\mathbf k}} \cdot\overline{ \mathcal M_{z,j,\mathbf k}}  e^{-W^{2}k^{2}/4} \sin 2\theta\nonumber\\
&&\times\left[\frac{e^{-i\omega_{j\mathbf{k}}t}-1}{\omega_{j\mathbf{k}}}-\frac{e^{-i\omega_{j\mathbf{k}}t}-e^{-ic_{l}kt}}{2(\omega_{j\mathbf{k}}-c_{l}k)}-\frac{e^{-i\omega_{j\mathbf{k}}t}-e^{ic_{l}kt}}{2(\omega_{j\mathbf{k}}+c_{l}k)}\right].\nonumber\\
\label{mzRkl}
\eer
from the pressure. The angular dependences, $\cos 2\theta$ and $\sin 2\theta$, come from those in Eq.~(\ref{HT}).

We adopt the following parameters~\cite{Hashimoto17} for Lu$_{2.3}$Bi$_{0.7}$Fe$_{4.2}$Ga$_{0.8}$O$_{12}$ with saturation magnetization $\mu_{0}M_{0}={78\,}\mathrm{mT}$. The transverse and longitudinal sound velocities are $c_{t}=3000\,\mathrm{m/s}$ and $c_{l}=5500\,\mathrm{m/s}$, and the Gilbert damping coefficient is typically $10^{-4}$~\cite{Manuilov09}. The gyromagnetic ratio is $\gamma/2\pi=28\,\mathrm{GHz/T}$. The cubic anisotropy constant $K_c =230$~J/m$^3$ and uniaxial anisotropy constant $K_u=-1200$~J/m$^3$, supplying internal fields $2K_c/M_0=7.4$~mT and $2K_u/M_0=-38$~mT. The thickness $d=4$~$\mu$m. These parameters used in Eq.~(\ref{eigen_mode}) reproduce the measured spin wave dispersions~\cite{Hashimoto17}. Assuming $d\ll d_p$, $e^{-z/d_p}\approx 1$.

In Fig.~\ref{Mechanisms} we plot the snapshots at 4~ns after the pump pulse. The left and right panels illustrate the transient response to shear and pressure stresses, respectively. The results for different applied magnetic fields and spot sizes display common features such as two(four)-node circular wave fronts from the shear(pressure) actuation. The diameter of the wave front for shear (pressure) stress is 12 (22)~$\mu$m, which agrees with  $c_t t$ ($c_l t$), i.e., the propagating distance of the transverse (longitudinal) acoustic pulses from the origin in a time interval $t$.  

\begin{figure}[ptb]
\includegraphics[width=8.5cm]{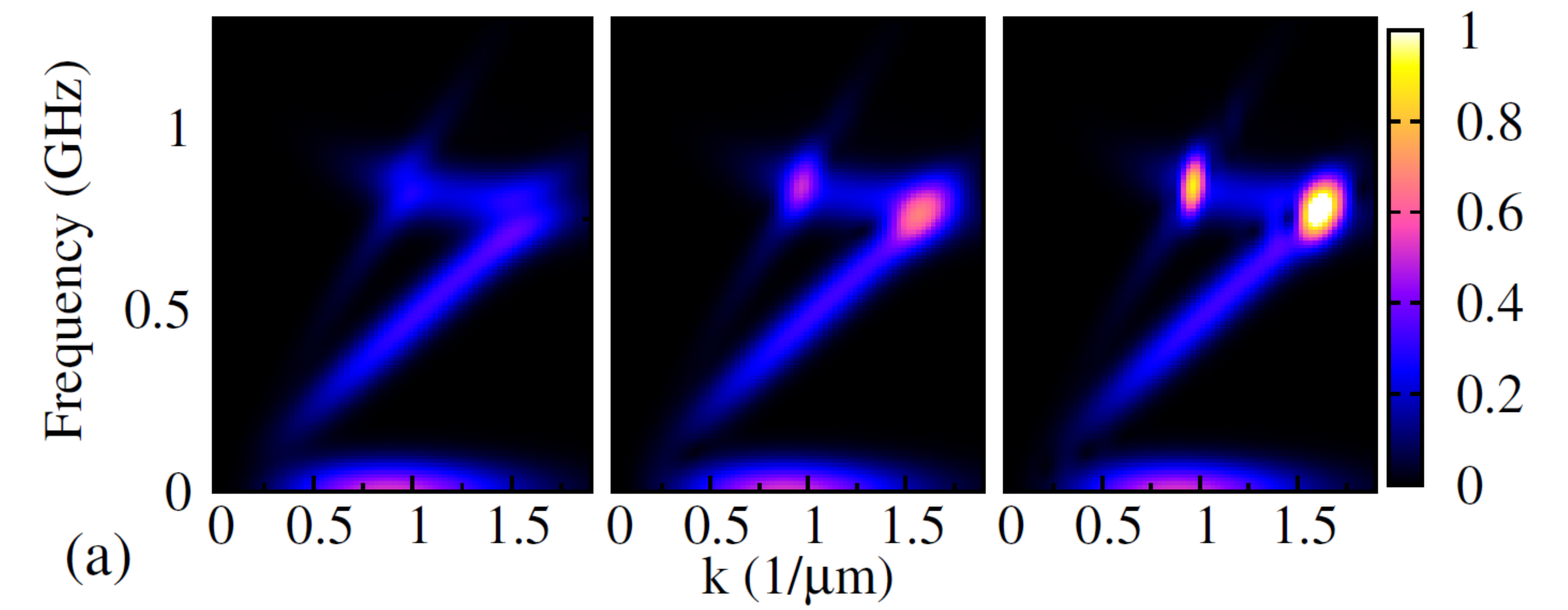}
\includegraphics[width=8.5cm]{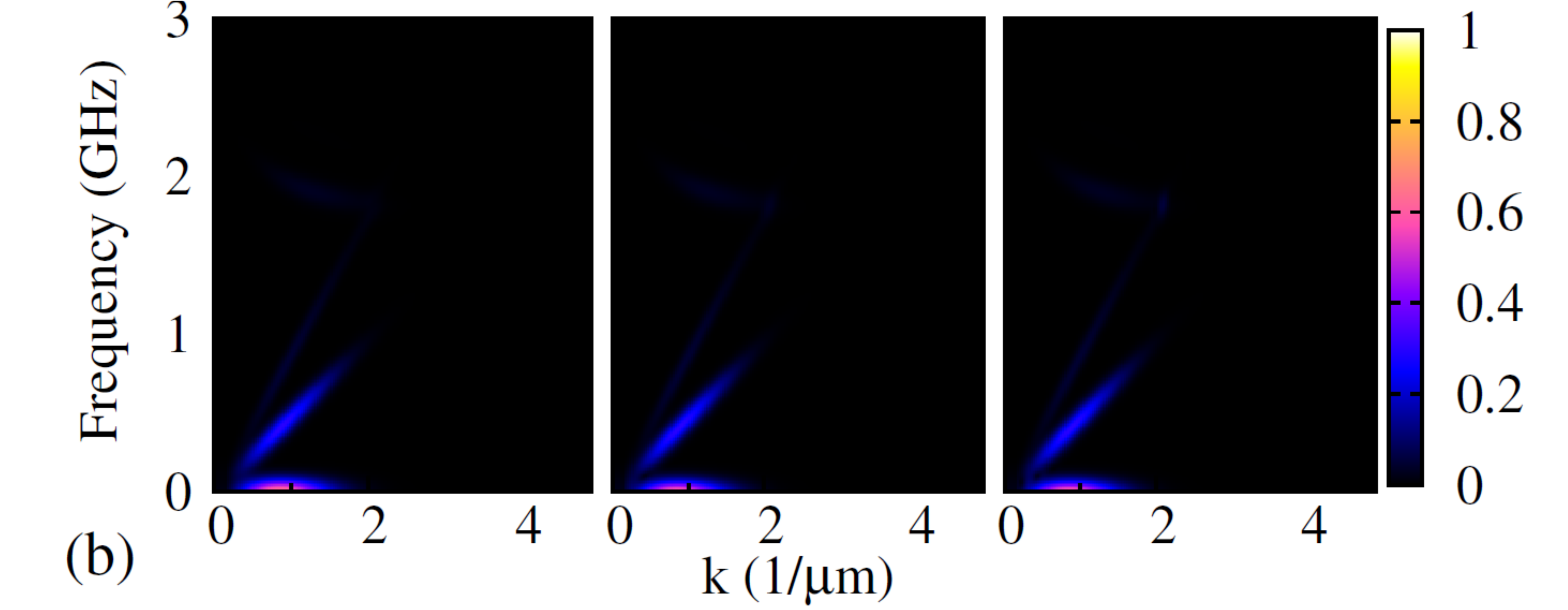}
\includegraphics[width=8.5cm]{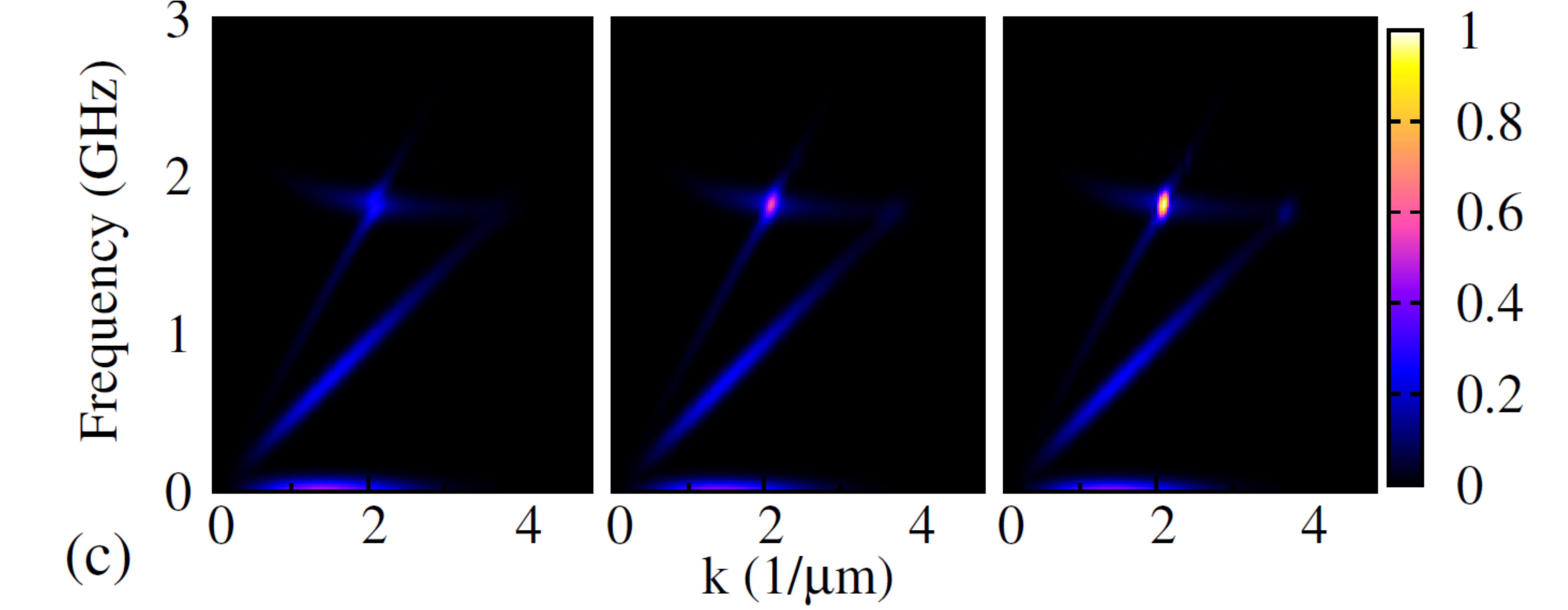}
\caption{Gabor spectra at center times $\tau$=2, 5, and 10~ns, for left, middle and right panels, respectively. Here $\theta=\pi/12$ and the Gabor window $\tau_d=4$~ns. The spot size and external field in the upper panels are 2.5~$\mu$m and 4~mT, while those in the lower (middle) panels are 1.5 (2.5)~$\mu$m and 40~mT, respectively.}%
\label{WK_Gabor}%
\end{figure}

Figs.~\ref{Mechanisms}(a) and (b) are calculated for a magnetic field $\mu_0H=4$~mT and spot size $W=2.5$~$\mu$m. In addition to the wave front we observe additional fringes due to the strong coupling in reciprocal space at the (anti)crossing point of the spin wave and phonon dispersions~\cite{Shen15,Kikkawa_polaron,Kittel58}. The Gaussian factor $e^{-k^2W^2/4}$ in Eqs.~(\ref{mzRkt}) and (\ref{mzRkl}) suppresses spin waves with $k>k_c= 2/W$. At 4~mT, the wave vector of the intersection point $k_{in}$ of spin wave and longitudinal (transverse) phonon branches is around 0.9(1.7)/$\mu$m [see Fig.~\ref{WK_Gabor}(a)], which is comparable with $k_c$, implying the excitation of magnon-polarons, i.e., the fully hybridized state of the magnons and phonons~\cite{Shen15,Kikkawa_polaron}. Since the longitudinal component (along $\mathbf k$) of the spin wave group velocity is smaller than the sound velocities, the spin waves lag behind the acoustic pulse and generate the observed ripples. 
The wavelength of these ripples reflects the wavelength of the spin waves at the crossing [also see Fig.~\ref{damping}(b)].
With increasing magnetic field, the spin wave dispersion is shifted to higher frequencies and the intersection points move to larger momenta, such that ultimately $k_c< k_{in}$. In this limit $\omega_{j\mathbf k}\pm c_t k\sim \omega_{j\mathbf k}$ in Eqs.~(\ref{mzRkt}) and (\ref{mzRkl}), which cancels the spin wave phase $e^{-i\omega_{j\mathbf k}t}$. The dynamics is then solely governed by the acoustic pulse. Indeed, in Fig.~\ref{Mechanisms}(c) and (d) for 40~mT the primary circles around the wave front dominate. In Fig.~\ref{Mechanisms}(e) and (f), we shrink the size of the laser spot to 1.5~$\mu$m and by enlarging $k_c$ recover (somewhat) the fringes inside the ring in the pressure actuated signal in Fig.~\ref{Mechanisms}(f).

More information can be distilled by a  Gabor transformation of the spatiotemporal spectra $\Theta_F(\mathbf r,t)$ \cite{Hashimoto17}, i.e., by calculating the time-resolved spectra $\Theta^G_{F}(\omega_0,\tau)=(1/\tau_g \sqrt{\pi})\int dt e^{-i\omega_0 t}\Theta_{F}(t)\exp(-(t-\tau)^2/\tau_g^2)$. $\tau_g$ stands for the time window of the  Gabor transform and $\tau$ is the center time. For the shear stress response this leads to
\ber\Theta^G_{F}(\mathbf{k},\omega_0,\tau)	&\propto& 	\cos\theta\sum_{j}|\overline{{\cal M}_{z,j,\omega_0,\mathbf k}}|^2ke^{-W^{2}k^{2}/4}e^{i\omega_0\tau}\nonumber\\
&&\hspace{-1.5cm}	\times\Big(\frac{e^{-i\omega_{j\mathbf{k}}\tau-\tau_g^2(\omega_{j\mathbf{k}}-\omega_{0})^{2}/4}-e^{ic_{t}k\tau-\tau_g^2(\omega_{0}+c_{t}k)^{2}/4}}{2(\omega_{j\mathbf{k}}+c_{t}k)}\nonumber\\
&&\hspace{-1.5cm}+\frac{e^{-i\omega_{j\mathbf{k}}\tau-\tau_g^2(\omega_{j\mathbf{k}}-\omega_{0})^{2}/4}-e^{-ic_{t}k\tau-\tau_g^2(c_{t}k-\omega_{0})^{2}/4}}{2(\omega_{j\mathbf{k}}-c_{t}k)}\nonumber\\
&&\hspace{-1.5cm}-\frac{e^{-i\omega_{j\mathbf{k}}\tau-\tau_g^2(\omega_{j\mathbf{k}}-\omega_{0})^{2}/4}-e^{-\tau_g^2(\omega_{0})^{2}/4}}{\omega_{j\mathbf{k}}}\Big). \label{eq:Gabor}
\eer 
Assuming equal importance of shear and pressure stresses, we plot in Fig.~\ref{WK_Gabor} the total response (absolute value of $\Theta^G_F$) as function of frequency $\omega_0$ and the wave vector modulus $k$ along $\theta=\pi/12$ for three center times $\tau=2$ (left), 5 (middle), and 10~ns (right) after the arrival of the pump pulse. 

\begin{figure}[ptb]
\includegraphics[width=3.5cm]{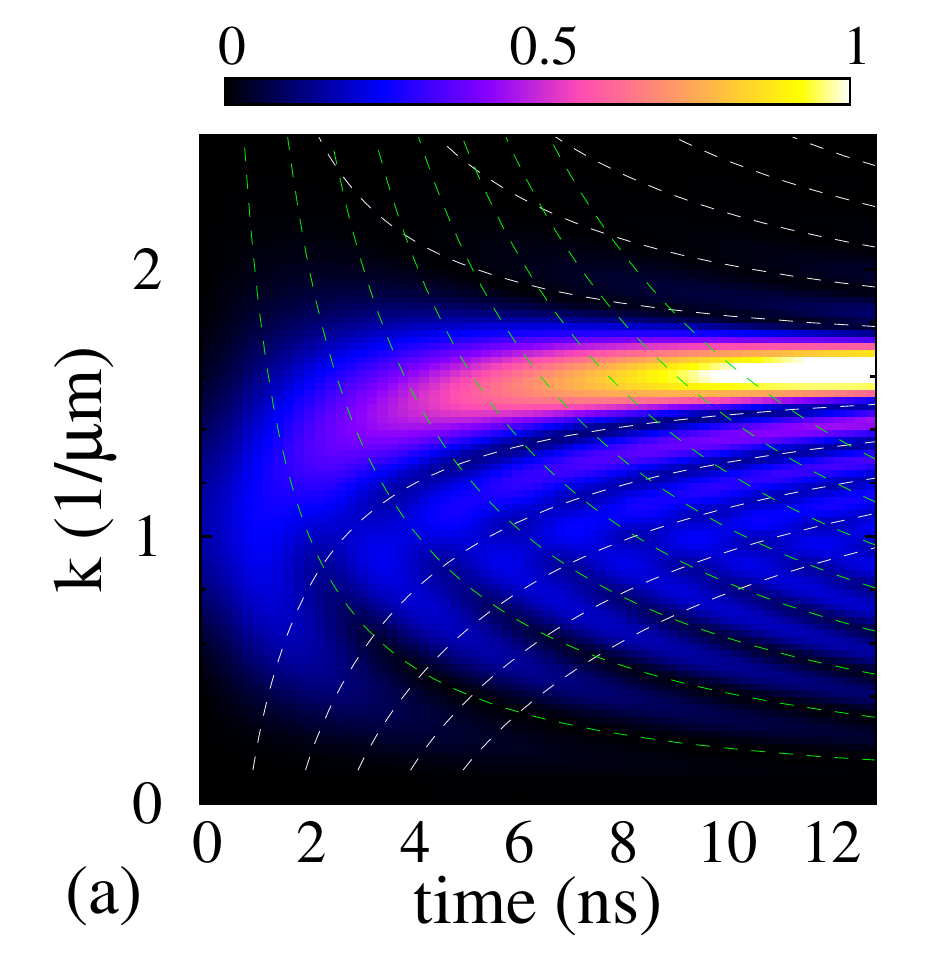}
\includegraphics[width=3.5cm]{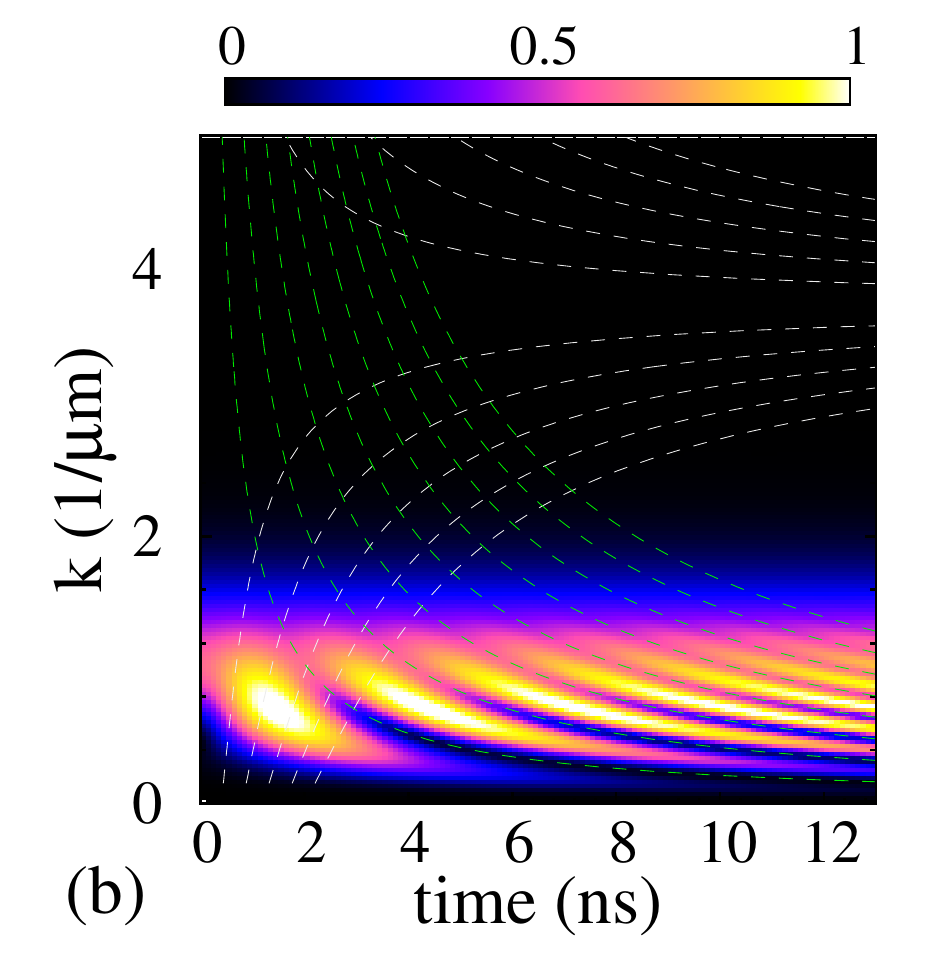}
\caption{(Color online) Time evolution of Gabor spectra along $\omega_0=c_t k$ for $\theta=\pi/12$ and time window $\tau_d=$1~ns with magnetic field (a) 4~mT and (b) 40~mT, respectively. The spot size is $W=2.5$~$\mu$m.
On the green and white dashed lines $c_t k\tau/(2\pi)$ and $(\omega_{j\mathbf k}-c_t k )\tau/(2\pi)$ become integers, respectively.}%
\label{KT_Gabor}%
\end{figure}

\begin{figure}[ptb]
\includegraphics[width=3.5cm]{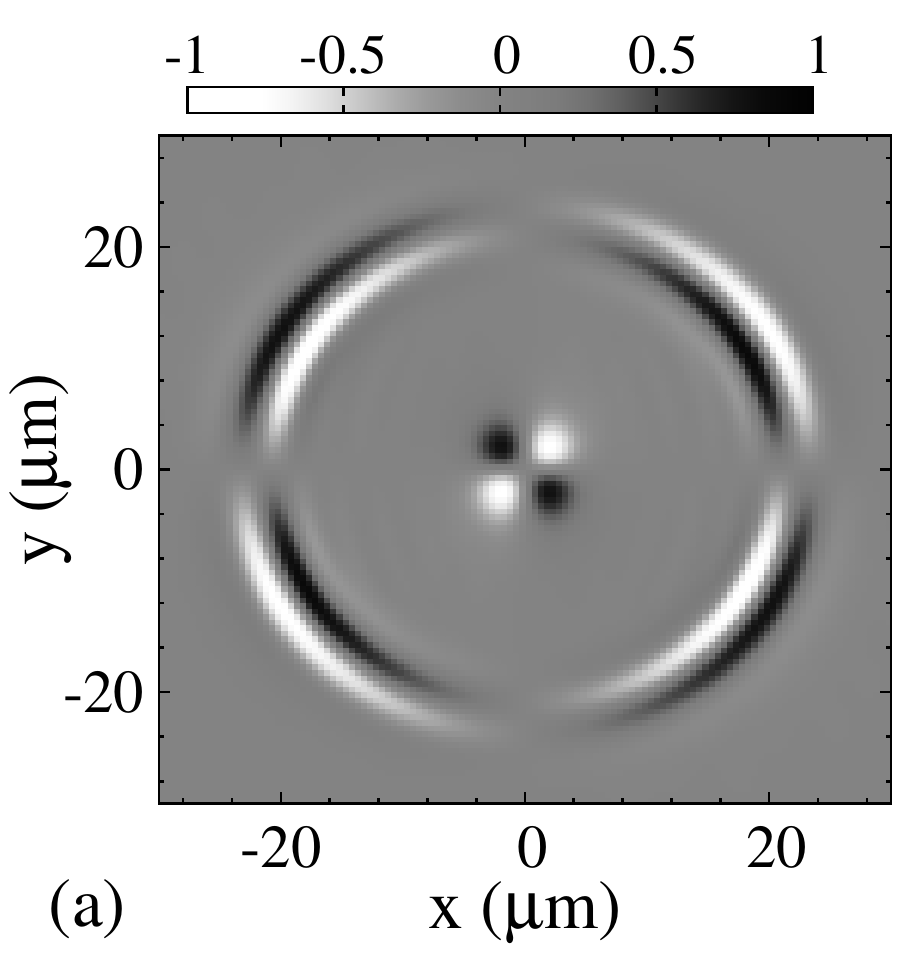}
\includegraphics[width=3.5cm]{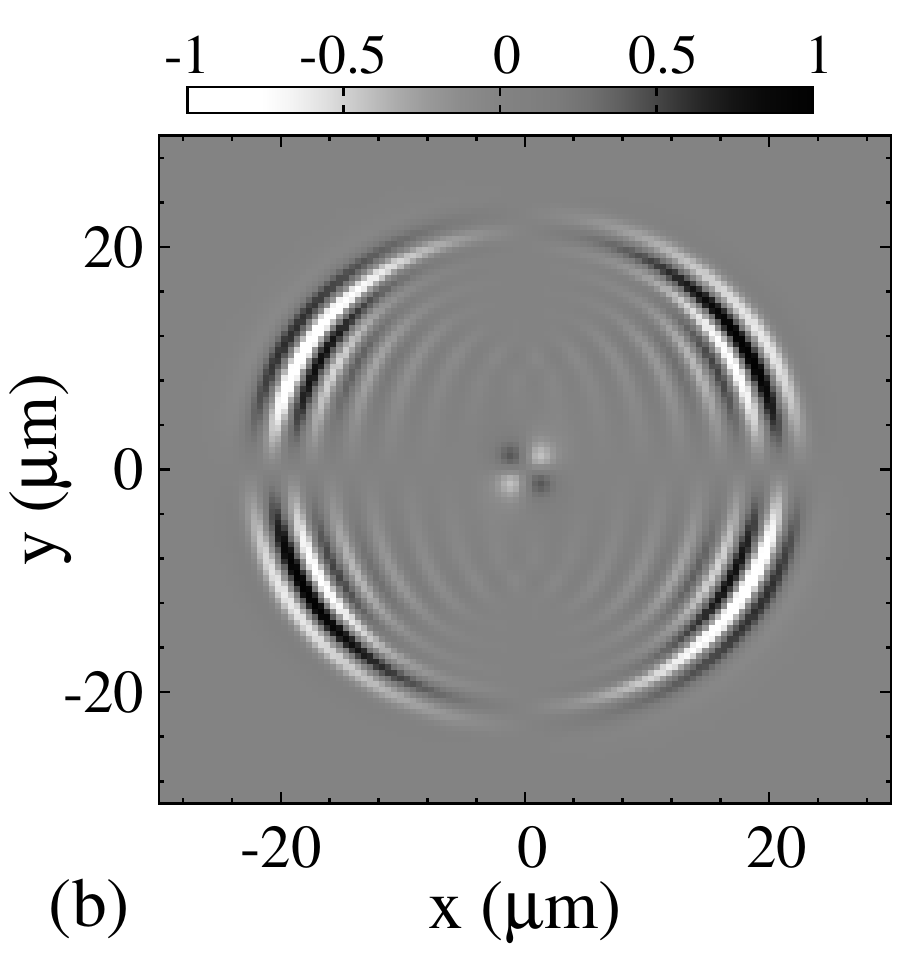}
\caption{Snapshots at 4~ns after ultrafast laser pulse due to pressure stress with external magnetic field at 40~mT. The spot size $W$ is taken to be (a) $2.5$~$\mu$m and (b) $1.5$~$\mu$m, respectively and the Gilbert damping is enhanced to $\alpha=0.1$.}%
\label{damping}%
%/Hashimoto/film/0for_paper/0with_anisotropy/demagnetization/osummary_mechanism.lyx
\end{figure}

The parameters in Fig.~\ref{WK_Gabor}(a) are the same as those in Fig.~\ref{Mechanisms}, i.e., $W=2.5$~$\mu$m and $\mu_0 H=4$~mT. The flat and linear bands correspond to spin wave and (transverse and longitudinal) acoustic wave dispersion relations, respectively. One may wonder why the pure phonon branches are so visible in a measurement of the magnetization. The answer can be traced back to the two phase factors, $e^{\pm ic_{t(l)}kt}$ and $e^{-i\omega_{j\mathbf{k}}t}$, in Eqs.~(\ref{mzRkt}) and (\ref{mzRkl}). The former is the phase of the AC driving field (from the phonons) while the latter is that of the responding oscillator (spin wave). When the frequency of the AC driving resonates with a spin wave mode the response is dominated by $(e^{-i\omega_{j\mathbf{k}}t}-e^{-i c_{t(l)}kt})/{(\omega_{j\mathbf{k}}-c_{t(l)}k)}\propto t$. The magnetization amplitude at the intersection of the spin waves with both transverse and longitudinal phonon dispersions is strongly enhanced and increases with time, reflecting the energy transfer from acoustic to spin waves under resonant condition [also see Fig.~\ref{KT_Gabor}(a)].
 Note that the divergence for large $t$ is cut-off by the damping and hybridization that is disregarded in Eq.~(\ref{eq:Gabor}). When the magnetic field increases to 40~mT in Fig.~\ref{WK_Gabor}(b), the resonance is shifted far out into momentum space and is not significantly excited anymore [also see Fig.~\ref{KT_Gabor}(b)]. 
The resonant excitation (by pressure stress) is switched on again in Fig.~\ref{WK_Gabor}(c) in which the spot size reduced to 1.5~$\mu$m, which is consistent with the interpretation of Fig.~1 above. The zero frequency branches in Fig.~\ref{WK_Gabor} originate from the time-independent term in the bracket of Eqs.~(\ref{mzRkt}) and (\ref{mzRkl}), which is the magnetic response to static local thermal expansion in the exposed area (see left panel in Figs.~\ref{Mechanisms} and Figs.~\ref{damping}).

The phase factors in Eq.~(\ref{eq:Gabor}) cause interference features in the time domain. Focusing on the phonon branches by substituting $\omega_0=c_t k$ and  $\tau_g\to 0$, 
\ber
|\Theta^G_{F}(\mathbf{k},\omega_0,\tau)|	&\sim& 
\Big|\frac{1-e^{i\omega_{j\mathbf k}\tau}}{\omega_{j\mathbf{k}}}-\frac{1-e^{i(\omega_{j\mathbf k}-c_{t}k)\tau}}{2(\omega_{j\mathbf{k}}-c_{t}k)}\nonumber\\
&&-\frac{1-e^{i(\omega_{j\mathbf k}+c_{t}k)\tau}}{2(\omega_{j\mathbf{k}}+c_{t}k)}\Big|  \label{dyn_Gabor}
\eer 
vanishes when $\omega_{j\mathbf k} \tau/(2\pi)$ and $(\omega_{j\mathbf k}-c_t k )\tau/(2\pi)$, and hence $c_t k\tau/(2\pi)$, are integers. For finite $\tau_g$, the cancellation is not exact, but as can be seen in Fig.~\ref{KT_Gabor} with $\tau_g=1$~ns, an interference pattern remains visible in the time evolution of the Gabor spectra that reflects the condition  for destructive interference derived from Eq.~(\ref{dyn_Gabor}) [see green dashed curves for integers of $c_t k\tau/(2\pi)$ and white dashed curves for $(\omega_{j\mathbf k}-c_t k )\tau/(2\pi)$]. These interference patterns have already been observed and reported in Ref~\cite{Hashimoto17}, but without the explanation provided here.

Interference effects are also visible in the real space dynamics such as, for example, the hot spots around $y=\pm 15$~$\mu$m in Figs.~\ref{Mechanisms} (b) (also seen in experiments~\cite{Hashimoto2017b}). The spin wave dispersion around the mode crossing depends weakly on $k$ but strongly on the angle $\theta$ between $\mathbf k$ and $x$-axis~\cite{Hashimoto17}. Although the longitudinal group velocity (along $\mathbf k$) is small, the transverse one (perpendicular to $\mathbf k$) can therefore be large, allowing spin waves to propagate sideways to the local wave front~\cite{Shen15}. When the magnetic damping is sufficiently weak, many spin waves with different group velocities live long enough to arrive at the same location and interfere with each other, resulting in the hot spots. In samples with large magnetic damping, these interference patterns are suppressed as demonstrated already in Ref.~\cite{Shen15} and confirmed by Fig.~\ref{damping}(a) and (b), where we use a larger Gilbert damping constant $\alpha=0.1$. 
In this case, only the spin waves excited within a relatively short time interval ($\sim$lifetime) before the measurement instant survive. Those spin waves are not able to propagate over a sufficient long distance to create an interference pattern.
The remaining oscillations within the main wave front are stronger in 4(b) than (a), which is caused by the difference between resonant (b) vs. non-resonant (a) excitation: In (b), the wavelength of the propagating spin wave shows a clear angular dependence, because the spin wave dispersion and therefore the dominant crossing modes strongly depend on $\theta$~\cite{Hashimoto17}, 
while the dynamics in Fig.~\ref{damping}~(a) is mainly governed by the isotropic sound waves, as discussed above. Moreover, the static spin response to the local thermal expansion in the exposure area becomes visible in both (a) and (b) (see the feature around the origin) and is not suppressed by magnetic damping~\cite{Shen15}.

\begin{figure}[ptb]
\includegraphics[width=3.5cm]{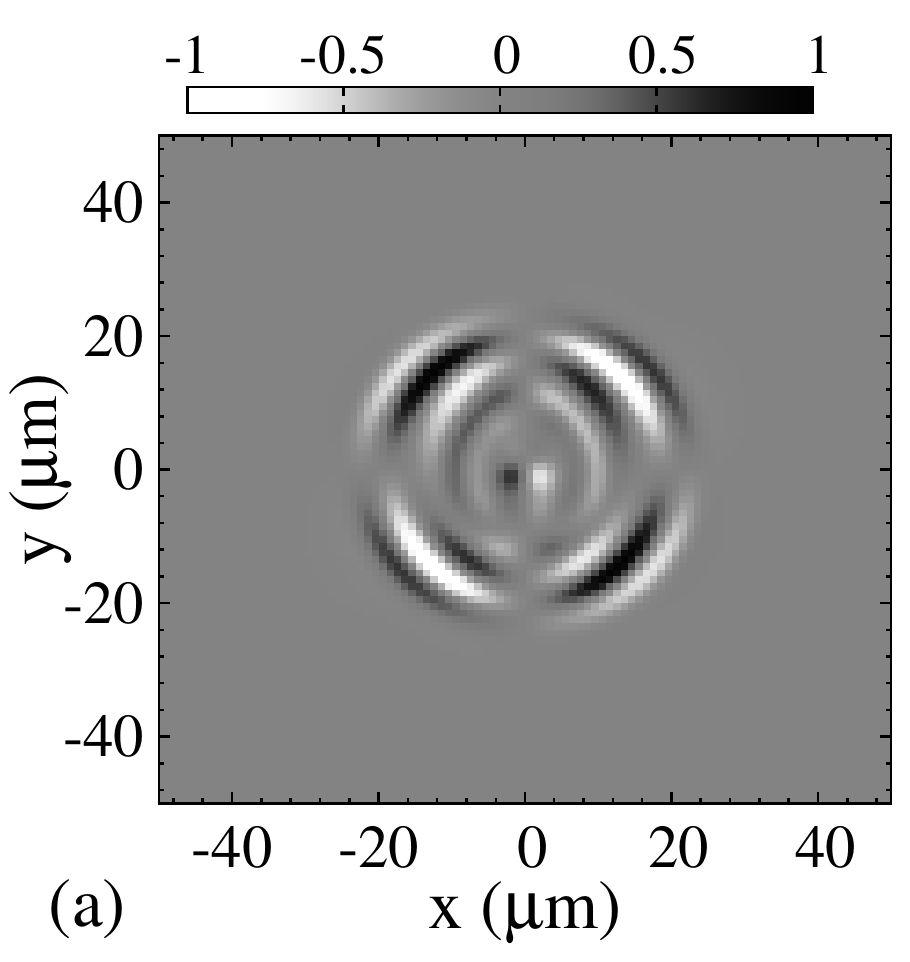}
\includegraphics[width=3.5cm]{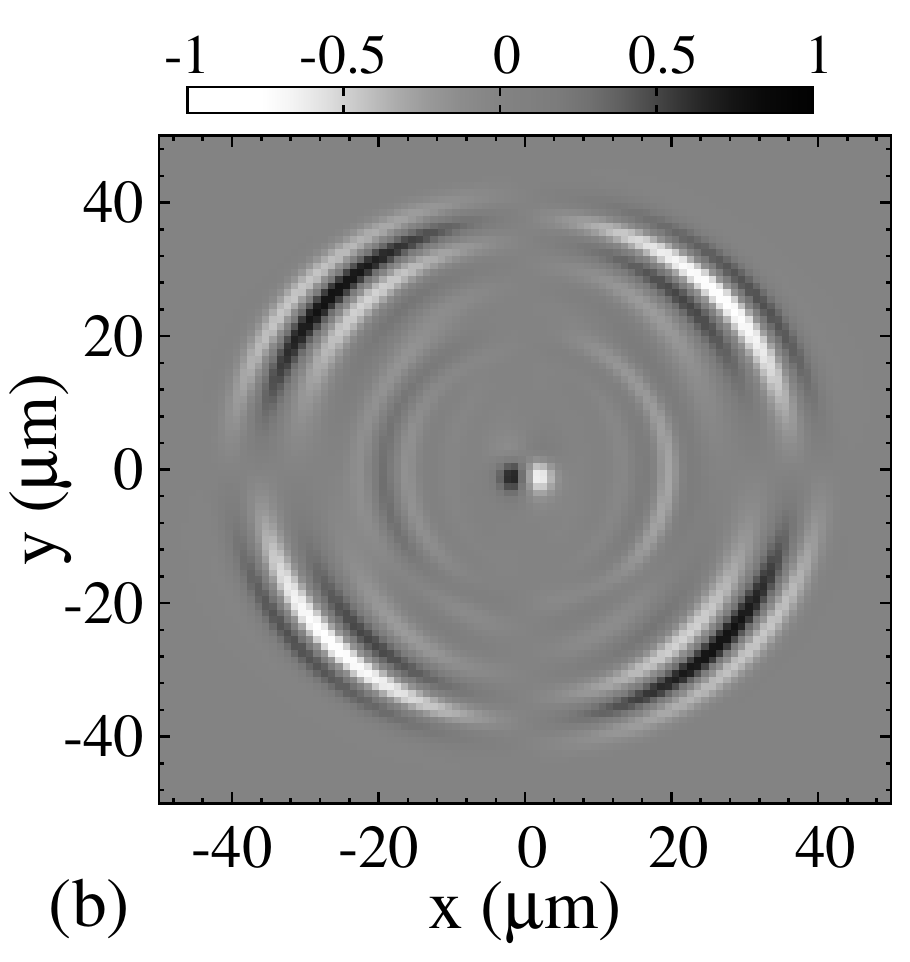}
\caption{Snapshots at (a) 3.45 and (b) 6~ns by fitting the experiment results in 40~$\mu$m thick doped iron Garnet by Ogawa et al.~\cite{Ogawa15}. The fitting parameters are $\alpha=0.25$, spot size $W=2.5$~$\mu$m, sound velocities $c_t=3.5$~km/s and $c_l=6.5$~km/s. In this calculation, $K_c=K_u=0$ are used.}%
\label{2Dfigure}%
\end{figure}

Finally, we apply our model to reanalyze the experiment by Ogawa et al.~\cite{Ogawa15}, where similar phenomena were found but explained by stimulated Raman scattering in the strong coupling regime. Our model shows that it may be caused solely by local heating. A close comparison with experiments requires the material parameters that are not all given in Ref.~\cite{Ogawa15}; we therefore adjust them to fit experimental data. We can reproduce most of the experimental curves with a single set of parameters in the caption of  Fig.~\ref{2Dfigure}, where we plot two snapshots for an external magnetic field of 70~mT. The good agreement with experiments supports our hypthesis that spin waves are generated by local heating. In the calculation we use a gyromagnetic $\gamma/2\pi=8.4$~GHz/T in order to fit the spin wave frequency at 70~mT. This might be caused by the fact that the magnetic anisotropy of the sample, is disregarded in our calculations by lack of better information. When we use the common value in insulator $\gamma/2\pi=28$~GHz/T, the internal magnetic field is estimated to be around 50~mT.

\section{Conclusion and discussion} 
In summary, we report calculations of the coupled lattice and  spin wave dynamics excited by focused laser pulses that agree well with recent experiments  by Hashimoto et al.~\cite{Hashimoto17,Hashimoto2017b}. The characteristics of resonant and non-resonant excitations, size effect and magnetic damping are included. Our model takes into account all dipolar-exchange spin wave modes. The model can be extended to include magneto-optical and transient demagnetization torques
generated by an ultrafast-laser pulse~\cite{Kimel05,Satoh12,Au13} by inverting our Eq.~(\ref{Dym}), although the present experiments 
analyzed here
do not provide unequivocal evidence for their existence. Very recently, Hashimoto et al.~\cite{Hashimoto2017c} interpreted the spatiotemporal distribution of the laser-induced magnetization dynamics in terms of the magnetic fields induced by demagnetization of the laser hot spot. Its symmetry is identical to that of a heat-induced pressure wave, so it is difficult to draw conclusions about the dominant excitation mechanism. Furthermore, the demagnetization cannot explain the patterns generated by the shear stress. We therefore conclude that the experiments by Ogawa et al.~\cite{Ogawa15} and Hashimoto et al.~\cite{Hashimoto17,Hashimoto2017b,Hashimoto2017c} are proof of magnetoelastic effects with possible contributions from heat-induced demagnetization. 

In principle, we can compute the energy flowing back from spin wave to acoustic wave by introducing the magnetoelastic coupling  non-perturbatively into the equations of motion for spin and acoustic waves as in Ref.~\cite{Shen15}, which we refrained from doing here, because the time scale of the magnetoelastic coupling strength is two orders magnitude larger than that of the spin wave dynamics ($\sim$ns). It is therefore a very good approximation to treat here the magnetoelastic coupling as a perturbation. A non-perturbative calculation would be necessary, for instance, to obtain the arcs in the left figure of Fig.~4(b) in Ref.~\cite{Shen15} which are caused by the inverse Faraday effect with large magnetic damping. The extension of the non-perturbative treatment to multi-spin wave modes is subject of a future study.

\begin{acknowledgements}
This work is supported by the DFG Priority Program 1538 SpinCat, the Netherlands Organisation for Scientific Research (NWO), and the JSPS (Grant Nos. 26103006). K.S. acknowledges the Recruitment Program of Global Youth Experts. 
\end{acknowledgements}
\bibliographystyle{prsty}
\bibliography{Refs.bib}
\end{document}